%% file: article.tex
\documentclass[12pt]{article}
\usepackage{epsfig}

\input econfmacros-csqcd3.tex
\def\Title#1{\begin{center} {\Large {\bf #1} } \end{center}}

\begin{document}

\Title{The Evolution of Proto-Strange Stars}

\bigskip\bigskip


\begin{raggedright}

{\it Omar G. Benvenuto\index{Benvenuto, O. G.}\\
Facultad de Ciencias Astron\'omicas y Geof\'{\i}sicas,
Universidad Nacional de La Plata,
Instituto de Astrof\'{\i}sica de La Plata (IALP-CONICET)\\
Paseo del Bosque S/N,
1900 La Plata,
Argentina\\
{\tt Email: obenvenu@fcaglp.unlp.edu.ar}\\
and\\
\it Jorge E. Horvath\index{Horvath, J. E.}\\
Instituto de Astronomia, Geof\'isica e Ci\^encias Atmosf\'ericas,
Universidade de S\~ao Paulo,
05570-010 Cidade Universit\'aria,
S\~ao Paulo, SP,
Brazil\\
{\tt Email: foton@iag.usp.br}}
\bigskip\bigskip
\end{raggedright}

\section{Introduction}

Since the idea that strange quark matter (SQM) may be the ground state of
hadronic matter \cite{Witten:1984rs} it has been a topic of great interest to
find ways to discern between neutron stars (NSs) and strange stars (SSs).
There have been several proposals related to this idea. For example, if the
transition from nuclear matter to SQM occurs during the cooling of a spin-down
pulsar it may be detected as a giant glitch \cite{Glendenning:1997fy} because
the star should undergo a sudden change in it moment of inertia. Another
possibility is to look for differences in the cooling of young presumed NSs
\cite{Schaab:1997hx}. Also, if it were possible to measure the radius of low
mass objects, we could distinguish SSs from NSs because of their smaller
radius \cite{Alcock:1986hz}.

If SQM actually were the ground state of matter, several interesting astrophysical
consequences should be expected. For example, there may exist high density cores
inside otherwise standard white dwarf stars. If these stars undergo non-radial
pulsation, the modes would be splitted on several close, detectable periods
\cite{Benvenuto:2005xs}. SQM formation may be the process that releases enough
energy to produce the core collapse supernova explosions \cite{Benvenuto:1989qr}.

In the conditions present in the first seconds after core bounce, the hadronic
matter that made up a {\it proto}-NS (PNS) contains a gas of degenerate electron
neutrinos. At the very beginning the evolution of the PNS is dominated by the
release of its neutrino content. It has been found that a gas of degenerate neutrinos
pushes away the critical density for the transition to quark matter. Thus, PNS
deleptonization {\it favors} the occurrence of the transition \cite{Benvenuto:1999uk}.
During the first minutes of the evolution of PNSs the thermodynamic conditions at
their interiors change so strongly that we think this to be the best place to detect
the transition to SQM. It should be remarked that supernovae light curves are
absolutely insensitive to the details of the explosion. So, the only way to observe
the transition to SQM (if it really occurs) is by observing the neutrino emission
of a forthcoming nearby core collapse supernova. If close enough, such an
event will allow the detection of a large number of neutrinos allowing for detailed
statistical studies. Unfortunately, the historical detections related to SN1987A
\cite{Hirata:1987hu} were not enough for this purpose because of the low number
of neutrinos.

In order to interpret future detailed neutrino observations it is relevant to have
available models of the behavior of NSs during its first seconds. This has been the
subject of several papers \cite{Burrows:1986me, Keil:1995hw, Pons:1998mm, Pons:2001ar}.
Here we shall present the first results we have found in the evolution of bare SSs.
This represents the starting point of an effort devoted to predict the details of
the neutrino signal due to the transition from nuclear matter to SQM. Here we do not
consider the occurrence of the transition to SQM during evolution, but center our
attention on the process of deleptonization of the SSs leaving the inclusion of
the physics of the quoted transition to future works.

The reminder of this paper is organized as follows: In Section \ref{relat_S_E}
we describe our General Relativistic, hydrostatic stellar evolution code.
In Section \ref{neu_opac} we briefly describe the physical ingredients we employed
and in Section \ref{results} we present our first results. Finally, in
Section \ref{final} we make some concluding remarks.

\section{Relativistic Stellar Evolution} \label{relat_S_E}

We have adapted our Newtonian stellar evolution code \cite{Benvenuto:2003pt} to
solve the equations of General Relativistic, hydrostatic stellar evolution
\cite{Pons:1998mm} in the diffusion approximation. The structure evolves with a
timescale far larger than transport, thus we solve it with two coupled Henyey
(finite differences, fully implicit) schemes: one for the structure and the other
for the transport \cite{Pons:1998mm}. Initially, at the  stellar interior,
neutrino mean free paths are far shorter than the size of the star. However this
is not the case at the outer layers or at later times. So, we adopted a flux
limiter to assure that causality is fulfilled.

The fluxes of lepton number $H_{\nu}$ and energy $F_{\nu}$ are

\begin{equation}  \nonumber
H_{\nu}= - \frac{T^2 e^{-\Lambda-\phi}}{6\pi^2}  \bigg[ D_{2} \frac{
\partial(T e^{-\phi})}{\partial r} +
(T e^{-\phi}) D_{3} \frac{\partial}{\partial r} \bigg(\frac{\mu_{e}}{T}\bigg) \bigg],
\end{equation}

\begin{equation}  \nonumber
F_{\nu}= - \frac{T^3 e^{-\Lambda-\phi}}{6\pi^2}  \bigg[ D_{3} \frac{
\partial(T e^{-\phi})}{\partial r} +
(T e^{-\phi}) D_{4} \frac{\partial}{\partial r} \bigg(\frac{\mu_{e}}{T}\bigg) \bigg]
\end{equation}

$T$ is the temperature, $\Lambda$ and $\phi$ are factors appearing in the
spherical Schwarszchild metrics, $r$ is the radius and $\mu_{e}$ is the electron
neutrino chemical potential. We consider electron, muon and tau neutrinos. In
general $\mu_e\neq0$ but muon and tau neutrinos are due to pair creation,
thus $\mu_\nu=0$, $\mu_\tau=0$. Then, diffusion coefficients are
$D_{2}= D_{2}^{\nu_{e}} + D_{2}^{\bar{\nu}_{e}}$,
$D_{3}= D_{3}^{\nu_{e}} - D_{3}^{\bar{\nu}_{e}}$, $D_{4}= D_{4}^{\nu_{e}} + D_{4}^{\bar{\nu}_{e}} + 4 D_{4}^{\nu_{\mu}}$ where

\begin{equation}  \nonumber
D_{n}^{j}= \int_{0}^{\infty} dx x^n f(E_{1}) \;
\frac{\big(1-f(E_{1})\big)^{2}}
     {\sum_{i}(\sigma_{i}/V)}, \;  \; \;j= \nu_{e}, \bar{\nu}_{e}, \nu_{\mu}.
\end{equation}

$\sigma_{i}$ represent the cross section of the reactions that provide the
neutrino opacity (see below) and $f(E_{1})=(1+\exp{\big[(E_1-\mu_1)/T\big]}$
is the occupation number corresponding to the considered neutrino. The equation
of lepton number per baryon $Y_{L}$ and energy conservation are

\begin{equation}  \nonumber
\frac{\partial Y_L}{\partial t} + e^{-\phi} \frac{\partial}{\partial a} \big( 4 \pi r^2 e^{\phi}F_{\nu} \big)= 0,
\end{equation}

\begin{equation}  \nonumber
e^{\phi} T \frac{\partial s}{\partial t} + e^{\phi} \mu_{e} \frac{\partial Y_L}{\partial t} +
e^{-\phi}\frac{\partial}{\partial a} \big( 4 \pi r^2 e^{2\phi}H_{\nu} \big)= 0.
\end{equation}

$s$ is the entropy per baryon and $t$ is the time. The equations of hydrostatic equilibrium, gravitational mass, radius, and metrics are

\begin{eqnarray}
\nonumber \frac{\partial P}{\partial a}= - \frac{e^\Lambda}{ 4 \pi r^4 n_B} ( \rho + P ) ( m + 4 \pi r^3 P ), \\
\nonumber \frac{\partial m}{\partial a} = \frac{\rho}{n_B e^\Lambda}, \\
\nonumber \frac{\partial r}{\partial a} = \frac{1}{ 4 \pi r^2 e^{\Lambda} n_B}, \\
\nonumber \frac{\partial \phi}{\partial a} = \frac{e^\Lambda}{ 4 \pi r^4 n_B}  ( m + 4 \pi r^3 P ).
\end{eqnarray}

$a$ represents the baryon number enclosed by a sphere of radius $r$ which is an adequate
 Lagrangian coordinate for our purposes, $P$ is the pressure and $n_B$ is the baryon
number density. At the center we have $r(0)=0$; $m(0)=0$;
$H_{\nu}(0)=0$; $F_{\nu}(0)=0$ whereas, at the surface
$\phi(a_s)= \frac{1}{2} \log{\big[2 m(a_s)/r(a_s)\big]}$ and $P(a_s)= P_s$.
The neutrino luminosity is $L_\nu= e^{2\phi} 4 \pi r^2 H_\nu$.

\section{Physical Ingredients}

\subsection{Neutrino Opacity} \label{neu_opac}

Neutrino opacity has been computed following the formalism presented in
Ref.~\cite{Steiner:2001rp}. As stated above, we considered electron,
muon and tau neutrinos assuming that $\mu_e\neq0$, $\mu_\nu=0$, $\mu_\tau=0$.
We considered the cross section per unit volume given by

\begin{eqnarray} \nonumber
\frac{\sigma}{V}=g \int \frac{d^3 p_2}{(2 \pi)^3}
\int \frac{d^3 p_3}{(2 \pi)^3}
\int \frac{d^3 p_4}{(2 \pi)^3}~~W_{fi} f_2
(1-f_3)(1-f_4)
(2 \pi)^4  \\ \nonumber \delta^4(p_1+p_2-p_3-p_4),
\nonumber
\end{eqnarray}

where the matrix element $W_{fi}$ is

\begin{eqnarray} \nonumber
W_{fi}= \frac{G_F^2}{E_1 E_2 E_3 E_4}&\bigg[ &
({\cal V}+{\cal A})^2 (p_1 \cdot p_2) (p_3 \cdot p_4 ) \nonumber \\
&+& ({\cal V}-{\cal A})^2 (p_1 \cdot p_4 )(p_3 \cdot p_2 )  \nonumber \\
&-& ({\cal V}^2-{\cal A}^2)(p_1 \cdot p_3 )(p_4 \cdot p_2 ) \bigg]. \nonumber
\end{eqnarray}

$E_i$ and $p_i$ are the energy and momentum of each particle participating in the
reactions (e.g., $\nu_{e}+d\rightarrow e^{-}+u$, $i= 1, 2, 3, 4$ respectively).
$\cal V$ and $\cal A$ are coefficients corresponding to each reaction, given in
\cite{Steiner:2001rp}, and the other symbols have their standard meaning. We have
developed a Montecarlo integration scheme that provides the quark matter neutrino
opacity for the thermodynamic conditions relevant at SS interiors.

\subsection{Equation of State}

For the equation of state of the SQM we have adopted the standard description
provided by the MIT bag model. In the zero strange quark mass limit,
it is well known that the form of the equation is $P = {1\over{3}} (\rho - 4B)$,
with $B$ a constant that parametrizes the non-perturbative interactions
giving rise to confinement. In the simulations presented below we adopted
the ``standard'' value of $B= 60 MeV fm^{-3}$. When a finite value for the
quark mass is employed there are deviations from the simple form given above,
although the linearity still holds to a high degree.

\section{Results}  \label{results}

\begin{figure}[htb]
\begin{center}
\epsfig{file=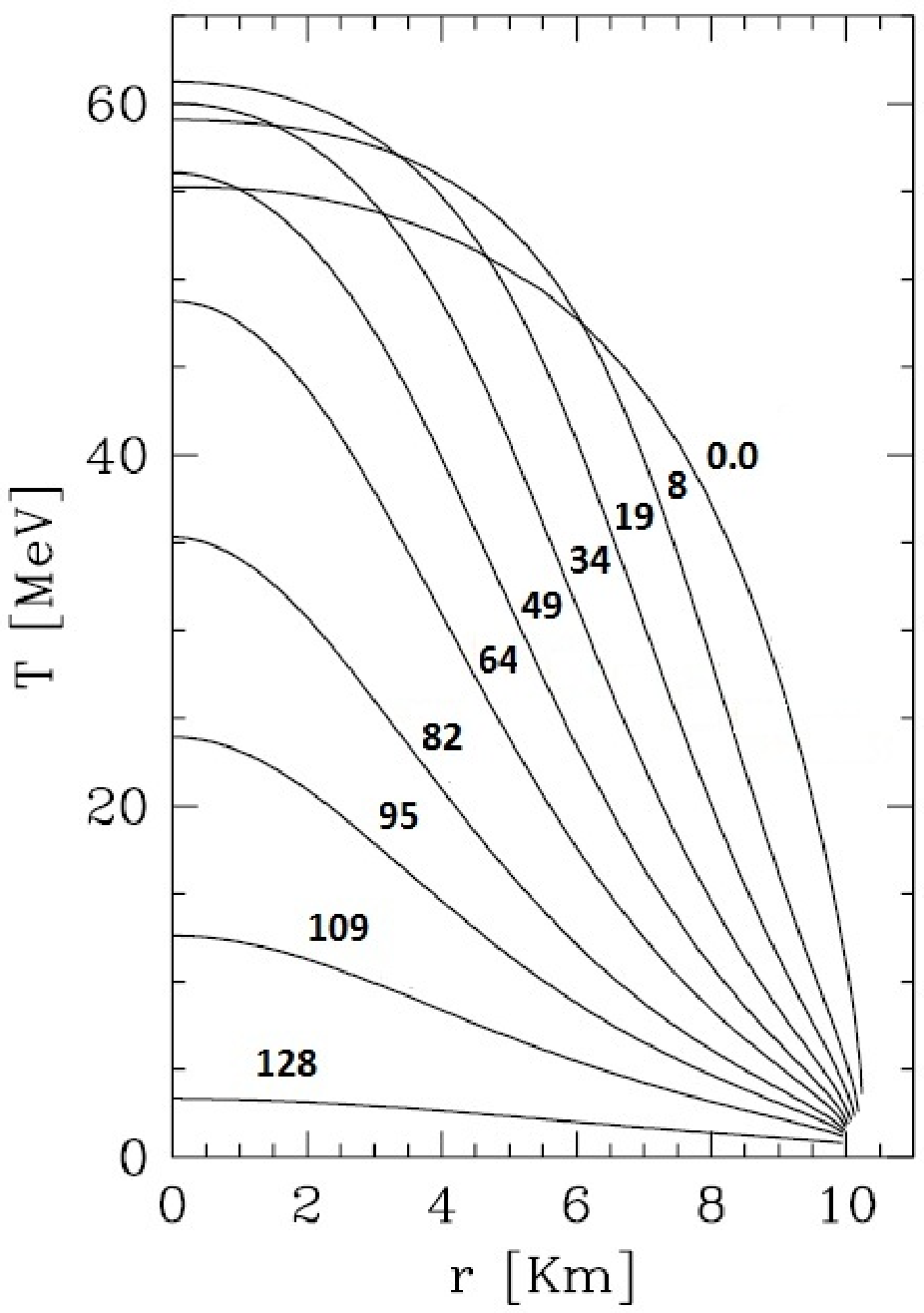,height=3.5in} \epsfig{file=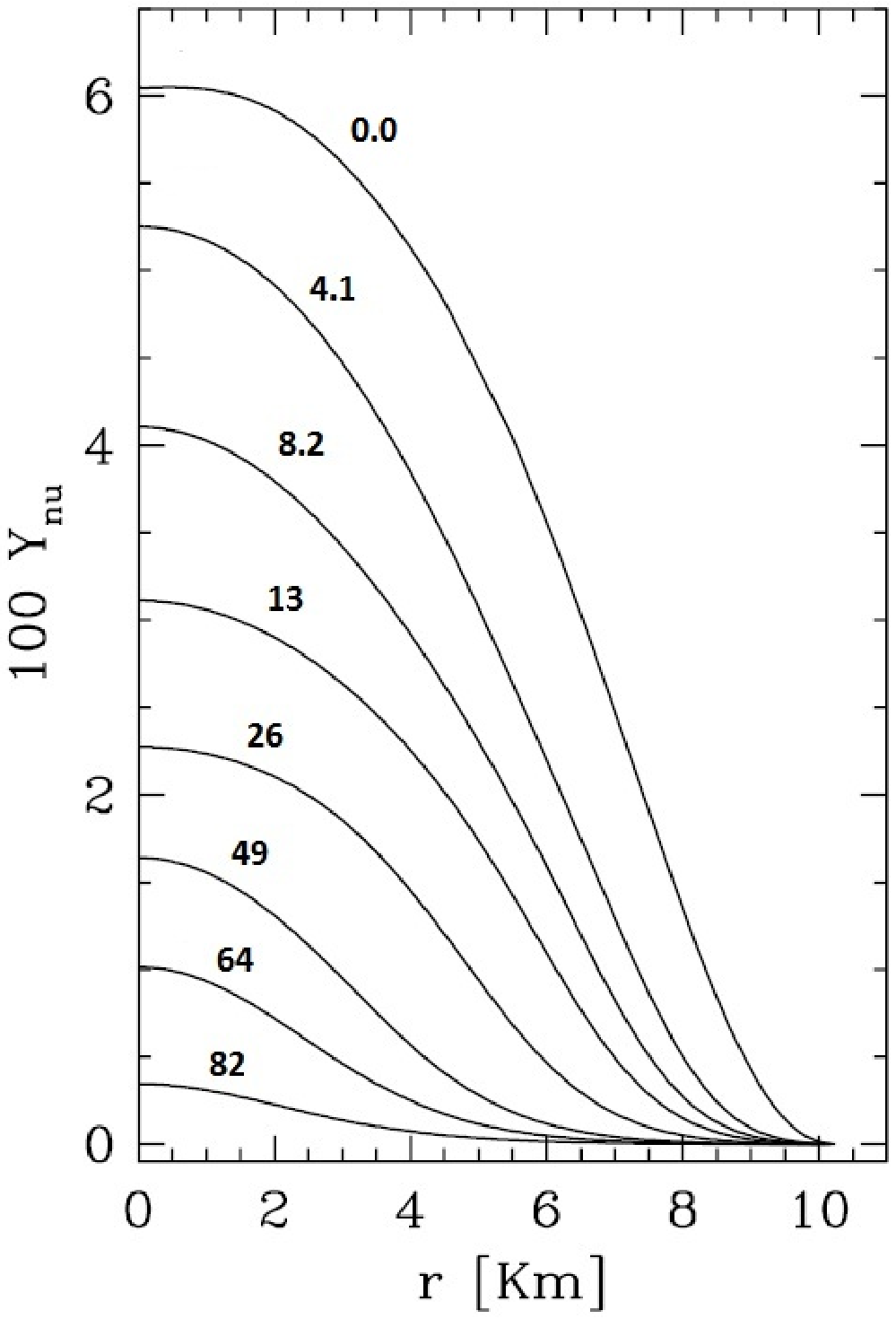,height=3.5in}
\caption{Left panel: The temperature profile during the first seconds of
evolution of a 1.4~M$_\odot$ homogeneous SS. Each curve is labeled with its age,
given in seconds on the right. Notice that, due to the outgoing neutrino flux at the fist stages
of evolution the inner layers of the star get hotter (Joule effect). Right
panel: The neutrino per baryon profile of the same SS. Notice that the timescale
of deleptonization is appreciably shorter than that of cooling.}
\label{fig:flux}
\end{center}
\end{figure}

We have applied our new relativistic stellar code to the case of a
1.4~M$_\odot$ homogeneous proto SS. We considered an initial energy
content compatible with that expected for a gravitational collapse of a
massive star. The evolution of the temperature and the neutrino abundance
per baryon of this object is depicted in Fig.~\ref{fig:flux}.

The proto SS is initially hot and plenty of neutrinos that tend to diffuse
outwards on a timescale of tens of seconds. The net leptonic number carried
by neutrinos is lost in the process termed ``deleptonization'' in the literature.
Heat (mainly in neutrino pairs) takes a much longer timescale, as indicated
by the comparison of both panels of Fig.~1. Because the SS interior is strongly
degenerate, the star also undergoes a tiny contraction. In fact the gravitational
mass and radius evolve less than in the case of NSs, in which the outer layers
are partially degenerate and more sensitive to thermal effects.

In is interesting to compare these results with the Fig.~9 of Ref.~\cite{Pons:1998mm}.
Our calculations indicate that NSs undergo a much faster evolution as compared to SSs.
Certainly, this is important in predicting the neutrino signal of the phase transition
to SQM and also in interpreting the observed signal in SN1987A within these models.

\section{Conclusions} \label{final}

In this conference we have presented the first results of the evolution of a 1D, bare,
proto SSs.  This represents a starting point in our effort devoted to predict the
neutrino signal to be expected to arrive from the next nearby core collapse supernova.
Recent work has explored 2D simulations in which the acceleration of the conversion front 
on its way outwards could be addressed \cite{Roepke}, although several open questions 
remain in both type of modeling and their eventual consistency.
We have performed a detailed computation of the neutrino opacity and coupled it to a flux
limited, hydrostatic, General Relativistic stellar evolution code. Then, we applied
the code to the evolution of a 1.4~M$_\odot$ homogeneous SS finding a slower evolution
as compared to standard PNS evolution. The net lepton number is lost faster than the
thermal energy, and it may be expected that the signal in a neutrino detector should last
longer. However, a closer inspection to the Fig.~1 also shows that the temperature
of the neutrinosphere (the imaginary surface at which most of the neutrinos
freely escape) decreases a factor $\sim 2$ in the first $\sim 10 sec$, causing the mean
energy of the emitted neutrinos to fall below $\sim 5 MeV$ or so. Thus, the emission
of the leaking neutrinos stands, but they are less energetic and easily missed
unless the threshold of the detector is very low. This and other important features
should be studied in depth and a more detailed description of the results
is in preparation and will be published elsewhere.

\bigskip
O.G.B. deeply acknowledges the financial support he received from the LOC that allowed him to attend the meeting.

\end{document}

%% file: econfmacros-csqcd3.tex
\def\beq{\begin{equation}}
\def\eeq#1{\label{#1}\end{equation}}
\def\eeqn{\end{equation}}

\def\beqa{\begin{eqnarray}}
\def\eeqa#1{\label{#1}\end{eqnarray}}
\def\eeqan{\end{eqnarray}}

\let\bar=\overbar

\def\Dslash{\not{\hbox{\kern-4pt $D$}}}
\def\dslash{\not{\hbox{\kern-2pt $\del$}}}

\def\msb{{\bar{\ssstyle M \kern -1pt S}}}

\usepackage{fancyhdr,graphicx}